\documentclass[10pt,letterpaper]{article}

\usepackage[utf8]{inputenc}
\usepackage{pslatex}
\usepackage{float} %
\usepackage[margin=1in]{geometry}
\usepackage{setspace}
\usepackage[T1]{fontenc}
\usepackage[utf8]{inputenc}
\usepackage{palatino}

\PassOptionsToPackage{hyphens}{url}\usepackage{hyperref}

\usepackage{comment}

\usepackage{lineno}

\author{Seth Frey\\Department of Communication, University of California, Davis\\Brian Keegan\\Department of Information Science, University of Colorado Boulder\\Peter Krafft\\Information School, University of Washington}

\date{}

\usepackage[table,xcdraw]{xcolor}
\usepackage{booktabs} 
\usepackage[normalem]{ulem} 
\usepackage{makecell} 
\usepackage{framed}
\usepackage{enumitem}

\usepackage{pifont}
\usepackage{lscape}

\usepackage[ruled]{algorithm2e} 

\SetAlFnt{\small}
\SetAlCapFnt{\small}
\SetAlCapNameFnt{\small}
\SetAlCapHSkip{0pt}
\IncMargin{-\parindent}

\usepackage[framemethod=TikZ]{mdframed}
\newcounter{infobox}[section]
\renewcommand{\theinfobox}{\thesection.\arabic{infobox}}

\title{Designing for Participation and Change in Digital Institutions}

\def\sf#1{{\color{red}\textbf{[SF: #1]}}}
\def\pk#1{{\color{violet}\textbf{[PK: #1]}}}

\begin{document}

\maketitle

\begin{abstract}
Whether we recognize it or not, the Internet is rife with exciting and original institutional forms that are transforming social organization on and offline.  
Issues of governance in these Internet platforms and other digital institutions have posed a challenge for software engineers, many of whom have little exposure to the relevant history or theory of institutional design.  Here, we offer one useful framework with an aim to stimulate dialogue between computer scientists and political scientists.
The dominant guiding practices for the design of digital institutions to date in human-computer interaction, computer-supported cooperative work, and the tech industry at large have been an incentive-focused behavioral engineering paradigm, a collection of atheoretical approaches such as A/B-testing,  and incremental issue-driven software engineering.  
One institutional analysis framework that has been useful in the design of traditional institutions is the body of resource governance literature known as the ``Ostrom Workshop''.  
A key finding of this literature that has yet to be broadly incorporated in the design of many digital institutions is the importance of including participatory change process mechanisms in what is called a ``constitutional layer'' of institutional design---in other words, defining rules that allow and facilitate diverse stakeholder participation in the ongoing process of institutional design change. 
We explore to what extent this consideration is met or could be better met in three varied cases of digital institutions: cryptocurrencies, cannabis informatics, and amateur Minecraft server governance. 
Examining such highly varied cases allows us to demonstrate the broad relevance of constitutional layers in many different types of digital institutions.

\end{abstract}

\section{Introduction}












In April 2018, Mark Zuckerberg, founder and CEO of a major online social networking platform, testified to members of the United States Senate and House of Representatives on the role of privacy in his platform's design. This Congressional testimony was noteworthy, not only because it generated global coverage, trending topics, and remixable memes, but because it placed representatives of two of the world's most influential institutional experiments---Facebook and U.S. Congress---in the same frame. The ensuing frenzy of posturing, apprehensions, and socio-technical suppositions highlighted the need for more robust conversations about how large, complicated \textit{digital institutions} can and should be designed. 

Institutions are the rules, norms, constraints, routines, roles, and other structures that people use to organize their social interactions~\cite{North_Institutions_1991,Ostrom:2005tg}. Institutions operate through regulative (rule-setting, monitoring, sanctioning, \textit{etc.}), normative (obligation, expectation, morality, \textit{etc}), and cultural-cognitive (taken-for-granted, comprehensible, shared understanding, \textit{etc.}) mechanisms~\cite{Scott_Crafting_2013}. 
As digital institutions, whose rules and structures are encoded or enforced through software,  come to play an outsized role in mediating human affairs \cite{lessig1999code,boyd2012critical}, it becomes imperative to understand the considerations involved in their design and evolution.



Digital institutions present at least two unique challenges compared to non-digital forms. First, the inhuman rigidity, precision, and efficiency of code produces a new type of power differential between platform owners and users. Second, the barrier presented by the technical skill set that is needed to build digital systems has led to a small, homogeneous group of \textit{de facto} technocrats as the ruling class of the digital world. These twin challenges compound each other: a small homogeneous group  restricted in their world view will have limited ability to predict problematic misalignments between available services and the needs of a diverse user base. At the same time, the precision of software enables unprecedented levels of literalism in monitoring compliance to protocols, while its rigidity creates a fragility to unanticipated situations and malicious re-purposing. These characteristics of digital institutions exacerbate incongruities between the narrow intentions of designers and the unanticipated uses, misuses, and demands of users. The volume, variety, and velocity of incongruities within digital institutions motivates the need for rapid experimentation cycles that top-down institutional design paradigms struggle to accommodate. 





Within the relevant areas of academic computer science, the dominant theoretically-informed approach to the design of digital institutions is epitomized in the work of Kraut and Resnick~\cite{kraut_building_book_2012}.  This paradigm relies on an individualistic, incentive-based, and top-down engineering approach to digital institution design.
In the present work, we re-evaluate this dominant approach in light of the serious challenges digital institutions have confronted since 2016. 
Concentrating on the formal and informal norms and rules that complement a community technology, specifically the formal rules that define how  communities change, we draw upon the innovative---but largely sidelined---work of~\cite{kollock_managing_1996} as a starting point, which in turn relies heavily upon the participatory view of institution design advanced by political scientist Elinor `Lin' Ostrom.  Ostrom's work earned her a share of the 2009 Nobel Memorial Prize in Economic Sciences, recognizing her contributions to the science and design of common-pool resource management institutions.

Ostrom's research showed that participatory deliberation and decision-making are crucial for allowing institutions to adapt to local contexts and achieve greater sustainability than what is possible through centralized, top-down planning. 
By contrast, typical work in digital institution design, such as that of Kraut and Resnick, ``optimization'' approaches of A/B testing, or other engineering-oriented approaches, has focused on laying out the design space for digital institutions with negligible attention to \textit{who} can make design decisions about the institution and \textit{how} those decisions are adjudicated. 
Authors in the top-down engineering tradition---even in drawing upon Ostrom in building parts of their framework---see ``a moral imperative to create online communities that work well'', regardless of whether it is ``social engineering'' or ``paternalism''~\cite[p. 9]{resnick_introduction_2012}. 
In contrast, Ostrom recognizes that choices about how to scaffold democratic participation must be at the fore for designing institutions that are responsive and sustainable. 


In the remainder of this paper, we translate Ostrom's institutional design and analysis framework to the problems of contemporary digital institutional design. We summarize the ambitious and multidimensional work of Ostrom and her colleagues with a focus on the design of an institution's ``constitutional layer,'' the layer that defines how a system changes itself and who is formally defined in the change processes. 

Institutional change is possible to anticipate because, like all institutions, digital institutions inevitably change in response to a variety of factors---imitating other institutions' successes, adapting to changes in the availability of resources, and generating new rules and relationships~\cite{aldrich_organizations_1999,scott_organizations_2015}.  Constitutional layers help to facilitate healthy change by engaging diverse stakeholder perspectives and enabling institutions to better adapt to local contexts.
We use three case studies to illustrate how Ostrom's institutional design framework can be applied for the analysis of digital institutions, and to argue for the importance of constitutional design layers for creating more accountable and resilient digital institutions.

\section{Background: Supporting online collective action}

There is a relatively long tradition in the areas of human-computer interaction and computer-supported cooperative work interrogating how values are embedded in software and technological artifacts \cite{winner1980artifacts,friedman1996bias,friedman2006value}.  In close alignment with the related areas of infrastructure studies \cite{galloway2004protocol,denardis2009protocol,gillespie2014relevance,pasquale2015black,lustig2016algorithmic,musiani2016alternative} and critical platform studies \cite{gillespie2018custodians}, we build on this tradition by focusing on software and information technologies as components of institutional structures, rather than software and technology as user-facing consumer products.  

Designing a digital institution involves tasks such as defining the transfer and sharing of resources, how the user community will be built and sustained, and the action space of participations: what forms of interaction that the architecture and interface will support. During the rise of digital institutions over the last two decades, a handful of approaches to these questions have been pursued by practicing computer and information scientists. The pre-theoretical approach stereotypical of technology firms and digital platforms has been ``build it and they will come'', relying on the practical experience and precedent of previous platforms to guide design decisions as well as the intuitions and agility of managers to implement design choices. A related approach is to make design choices based on profit maximization or proxies such as ``user engagement'', in the extreme case without regard to factors such as community sustainability or health. Academic work oriented towards this goal focuses on atheoretical design methods such as A/B-testing for maximizing engagement and machine learning for behavioral prediction. But these pre-theoretical and atheoretical approaches are increasingly unable to stay ahead of the accumulating socio-technical strains that are consequences of basic design choices made at the foundations of these digital institutions. 

Although related areas have some traction, institutional design and analysis perspectives are rare in computer and information science and engineering (CISE) fields: there are only 13 results in the ACM digital library and 19 results in the IEEE digital library for ``institutional design'' and 7 and 28 respectively for ``institutional analysis''. The absence of institutional design and analysis frameworks in CISE research conversations is conspicuous given that CISE practitioners design and implement digital institutions that have profound influence over the social, psychological, political, and economic lives of billions of people. 

In this section, we outline the features of the prevailing paradigm of top-down digital institutional design, introduce the Ostrom Workshop's concept of constitutional layers as an important component for designing sustainable institutions, and summarize previous work using Ostrom's framework to analyze digital institutions.  

\subsection{Digital institutions}

While institutions are the rules, norms, constraints, routines, roles, and other structures that people use to organize their social interactions, digital institutions are institutions whose rules or structures are at least partially encoded or enforced through software.  
Modern institutions vary in the degree to which they are digitized; digitization is altering many historical institutions while also creating entirely new institutional forms \cite{dimaggio2001social}. Institutions like government departments use digital systems to implement or support existing legal or socio-technical infrastructure: a citizen submitting taxes online is participating in a digital component of a larger institutional framework, and the form of these digital structures creates affordances (such as new ways to submit taxes) and inequalities (such as through the ``digital divide'' \cite{hargittai2002second}). Social media platform like Twitter are more conspicuously digital because they existed first as user-facing digital products.
However, even Twitter is not a wholly digital institution: Twitter-the-platform is managed by engineers employed by Twitter-the-corporation that is embedded within layers of other non-digital institutions like publicly-traded corporations and employment law.

  \subsection{The engineering tradition of digital institution design}
The most prominent HCI engineering approach to \textit{Building Successful Online Communities (BSOC)}~\cite{kraut_building_book_2012} pioneered by Kraut, Resnick, Kiesler, Riedl, and their colleagues~\cite{beenan_using_2004,ren_applying_2007,farzan_commitment_2011,ren_attachment_2012} (hereafter, ``the engineering approach'') created a theoretical foundation for digital institution design. Fifteen years later, this has directly informed the designs of social media platforms like Facebook. Kraut and Resnick seek to ``identify a wide variety of levers of change, features of online communities that can be deliberately and strategically chosen''~\cite{kraut_building_book_2012}. The ``behavior change'' paradigm advanced by Kraut and Resnick treats the crafting of social systems as an engineering problem approachable by a central designer with unilateral control over the mechanisms of choice and influence 

\begin{table}[tb]
    \centering
    \begin{tabular}{p{1in}p{4in}}
        \toprule
        \textbf{Challenge} & \textbf{Example design claim} \\
        \midrule
        \textit{Motivations} & ``Requests from high-status people in the community lead to more contribution than anonymous requests or requests from low-status members.'' \\ \midrule
        \textit{Commitment} & ``Highlighting a community's purpose and successes... can translate members' commitment... into normative commitment to the community.'' \\ \midrule
        \textit{Newcomers} & ``Entry barriers for newcomers may cause those who join to be more committed to the group and contribute more to it.'' \\ \midrule
        \textit{Regulation} & ``Moderation decided by people who are members of the community, are impartial, and have limited or rotating power will be perceived as more legitimate and thus be more effective.'' \\ \midrule
        \textit{Founding} & ``Ambiguity of scope for the community creates opportunities for adjustment and member ownership.'' \\
        \bottomrule
    \end{tabular}
    \caption{Examples of design claims and their corresponding design challenges from~\cite{kraut_building_book_2012}.}
    \label{tab:kr_design_claims}
\end{table}

Within the engineering approach, the centralized planner-governor  decomposes effective communities into aggregations of socio-technical functionality for engineering different dimensions of desirable community member behavior. Kraut and Resnick explicitly describe eight ``levers of change'' that are social or technical configurations reflecting design decisions made by managers, designers, and members of online communities~\cite[p.6--8]{resnick_introduction_2012}. 
\begin{description}
    \item[Community structure.] \textit{How is the community organized?} Size, homogeneity, subgroup structure, and recruitment of members.
    \item[Content, tasks, and activities.] \textit{What kinds of activities does the community support?} Self-disclosed, imported, volunteered, professional content; interdependent vs. dependent tasks; social vs. immersive activities.
    \item[Selection, sorting, highlighting.] \textit{How can members find the information that is best for them?} Dividing community into spaces, highlighting good content, removing inappropriate content, and feeds or recommendations of relevant/related content.
    \item[External communication.] \textit{How can members communicate beyond the community?} Sharing content, migrating identities, and importing relationships embeds communities .
    \item[Feedback, rewards, sanctions.] \textit{How do community members receive feedback about their behavior?} Ratings, rewards, and sanctions provide informal or formal changes in status.
    \item[Governance.] \textit{How do online communities employ social roles, rules, and procedures to govern member behavior?} Handling newcomers and conflicts; rules for behavior by position; procedures for decision-making.
    \item[Access controls.] \textit{What controls can be imposed on its members?} Limits on membership and actions; selection and permissions of moderators.
    \item[Presentation and framing.] \textit{How does the community use examples to compare behavior?} Privileging vs. hiding bad behavior; emphasizing similarities to other communities.
\end{description}

This is a generative framework for enumerating the types of digital institutional design decisions that need to be made, but is systematically silent on \textit{who} and \textit{how} these decisions are made. In response to critiques that online communities are not easily designed or controlled, Kraut and Resnick concede that people ``cannot be shaped or programmed in the way physical materials or software can'' but offer  that ``online communities can be designed and managed to achieve the goals that their owners, managers, or members desire'' through a combination of social and technical configurations~\cite[p.6]{resnick_introduction_2012}. 
Typically, the default is implementing institutional designs without the time, cost, and risk of soliciting community members' input. Table~\ref{tab:kr_design_claims} provides representative examples of Kraut et al.'s~\cite{kraut_building_book_2012} design claims. 

Responding to the criticism about designing online communities as forms of ``social engineering,'' Kraut and Resnick offer a one-page exposition on the ``Morality of Design'' weighing institutional design as a mechanism to ``elicit individual behavior that benefits the community'' against creating online communities that ``make the communities more attractive for their members or more productive''~\cite[p.9]{resnick_introduction_2012}. Drawing on the concept of choice architectures introduced by Thaler and Sunstein's ``libertarian paternalism''~\cite{thaler_libertarian_2003,sunstein_libertarian_2003,thaler_nudge_2009}, Kraut and Resnick argue that  encouraging compliance with rules through psychological and economic incentives--- behavior change institutional mechanisms such as default design decisions---are sufficient to drive collective behavior within a  digital institutions toward a community's goals. This engineering approach is silent on participatory institutional design and has only a narrow conception of the importance of designing processes that include input from the members of the community affected by these design decisions. 

Our criticisms may seem unfair. In the chapter ``Regulating Behavior in Online Communities'' the authors do consider democratic recommendations like the need for community participation in rule-making alongside their more technocratic recommendations, and explicitly invoke Ostrom to do so. They offer a fair overview of the reported benefits of democratic institutions and participatory mechanisms, with direct mappings of Ostrom's prescriptions to their own. But despite their consideration of democratic means, their imagined audience are nevertheless administrators who are empowered to unilaterally implement designs, and their most prominent argument for fostering participation --- participatory mechanisms raise perceptions of legitimacy by increasing compliance --- is strictly instrumentalist. Because the legitimacy-increasing benefits of participation can be achieved with superficially participatory mechanisms,  suggestion boxes, ticket trackers, and non-binding polls, the  engineering tradition's instrumental argument for participation can only motivate a superficial commitment to it.

There is a pernicious slippage that happens in the space between Kraut and Resnick's arguments about the morality of design defaults and the morality of readers. Kraut and Resnick explicitly ``leave moral judgments---about which goals are worth designing for---to our readers''~\cite[p.9]{resnick_introduction_2012}. A (small-D) democrat might instead ask, ``Why aren't moral judgments about the design of the community left to the \textit{community members}?'' In Kraut and Resnick's framing, the moral judgments of the reader-as-designer---not the community members-as-designers---have ultimate authority. This is not a pedantic point:  throughout \textit{Building Successful Online Communities}, the community designer is assumed to be a singular actor who can act unilaterally to implement design choices, not as a collective making a legitimate decision (\textit{i.e.}, participatory democracy) or a governor representing its constituents' interests (\textit{i.e.}, representative democracy). Leaving moral judgments about design goals as an exercise for the reader is itself a default design decision that not only defines substantive democratic participation out of the framework, but left a decade of digital institution design to the Machiavellian morality of the market.  The instrumentalist approach to the tools of behavior change  arguably engendered the ``fake news'' crisis, as malicious agents chained each system's behavior change mechanisms to create new exploits with the power to threaten entire democracies beyond the digital realm. Facebook, Twitter, and other major subscribers to the engineering approach are only beginning to re-evaluate their commitments to it.

There are several potential rebuttals to our critique. The first is a matter of scope: the online community that Kraut and Resnick imagine themselves serving is a relatively small, possibly voluntary common-interest group, whose low stakes and modest scale make the ``benevolent dictatorship'' of a single designer/administrator a much lower-risk governance model~\cite{reagle_authorial_2007}. Perhaps Kraut and Resnick should not be held accountable for applications of the engineering approach beyond the scales for which it was imagined. But absent explicit scoping of their counsel to this subset of communities, they bear responsibility for its runaway adoption by the largest and most consequential of engineered digital institutions. The second rebuttal is one that the Ostrom Workshop's ecological grounding actually abets: where there are many communities, each using different structures to pursue similar goals, and where the costs of founding and migrating between communities is low (as is often the case in online communities), communities will voluntarily adapt their designs to reflect members' moral needs to secure their participation. Two of our three cases consider ecosystems that pit communities against each other, and find varying levels of effectiveness for the competitive mechanism for simulating subservience to users. The third rebuttal is anthropic: (1) the difficulty of designing truly participatory institutions and (2) the bias in case studies and other empirical work towards successes means that it is easier to advance the science of centrally engineered digital institutions, and design guidance cannot be generalized beyond cases which happen to have singular designers. Overcoming these two shortcomings of current participatory frameworks is a fundamental goal of the Ostrom Workshop.

In the remainder of the paper we will explore how ecological factors and proper constraints can counteract the failures of the engineering approach in certain circumstances. But it is imperative that we privilege more democratic design traditions, especially for the most influential digital institutions. While online communities may once have been fringe institutions, the propaganda crisis of the 2016 U.S. elections demonstrates how deeply embedded they have become within powerful geo-political systems. Conflicts inevitably invite questions about rule-making: ``who makes the rules?'' and ``how can rules be changed?'' The initial conditions on which Kraut and Resnick make their technocratic recommendations and justifications for designing digital institutions may be appropriate for new and small communities, but a phase transition happens somewhere in the course of digital institutions' growth where technocratic decision-making is unable to efficiently aggregate information, elicit representative preferences, or make legitimate decisions.

\subsection{The democratic tradition of digital institution design}
The Ostrom Workshop represents a more general and democratic view of institution design, one with  increasingly recognized potential to transform digital institutions~\cite{silberman_2016_ostrom,Pitt:2015bk}. This body of work has worked to unify the findings of a community of scientists and practitioners across anthropology, political science, economics, sociology, and other branches of the social sciences under  a common framework. Elinor Ostrom's book \textit{Governing the Commons}~\cite{Ostrom:1990ws} offers commonly-cited ``design principles'' ---  The Design Principles for Community-based Natural Resource Management \cite{Ostrom:1990ws,Cox:2010tb} --- that have been taken up in diverse settings like lobster fisheries, forest management communities, USENET forums, and ancient systems of pasture, turbary, and estovers. The Design Principles distill governance lessons from ecologically diverse institutions from around the world, some of which are centuries old, isolating elements of successful resource systems and providing criteria for diagnosing a community's resource or governance problems.  	They are the most well-known contribution of the Ostrom Workshop outside that community, and have been used by many to analyze of the above digital institutions.

\begin{description}
     \item[User boundaries.] Clear boundaries between legitimate users and nonusers must be clearly defined.
     \item[Resource boundaries.] Clear boundaries are present that define a resource system and separate it from the larger biophysical environment.
     \item[Congruence with local conditions.] Appropriation and provision rules are congruent with local social and environmental conditions.
     \item[Appropriation and provision.] The benefits obtained by users from a common-pool resource (CPR), as determined by appropriation rules, are proportional to the amount of inputs required in the form of labor, material, or money, as determined by provision rules.
     \item[Collective-choice arrangements.] Most individuals affected by the operational rules can participate in modifying the operational rules.
     \item[Monitoring users.] Designated monitors who are accountable to the users monitor the appropriation and provision levels of the users.
     \item[Monitoring the resource.] Designated monitors who are accountable to the users monitor the condition of the resource.
     \item[Graduated sanctions.] Appropriators who violate operational rules are likely to be assessed graduated sanctions (depending on the seriousness and the context of the offense) by other appropriators, by officials accountable to the appropriators, or by both.
     \item[Conflict-resolution mechanisms.] Appropriators and their officials have rapid access to low-cost local arenas to resolve conflicts among or between appropriators and officials.
     \item[Minimal recognition of rights to organize.] The rights of appropriators to devise their own institutions are not challenged by external governmental authorities.
     \item[Nested enterprises.] Appropriation, provision, monitoring, enforcement, conflict resolution, and governance activities are organized in multiple layers of nested enterprises.
 \end{description} 
 One classic success of the  framework was a large-scale comparative study of over 100 farmer- and state-managed irrigation works in rural Nepal, which revealed the increased ability of self-governing watersheds to respond to local material, environmental, and cultural conditions \cite{Tang:1992ty,ostrom1995incentives}. 

Although still not widely appreciated or accepted as standard canon in HCI or CSCW, the Ostrom Workshop has been productively applied in several cases in the HCI and CSCW literature that have taken an interest in self-government, governance and resource management, and engineering participatory online communities. Ever since Peter Kollock's analysis of USENET~\cite{kollock_managing_1996}, resource management perspectives have had a small foothold in the study of digital institutions.  Major contributions have been \textit{Knowledge Commons} by Hess and Ostrom~\cite{Hess:2007we} and \textit{Internet Success} by Schweik and English~\cite{Schweik:2012tw}. 
A whole literature has emerged around institutional analyses of the archetypal knowledge commons, Wikipedia~\cite{Forte:2009wl,Forte:2008eu,Heaberlin:2016fv,Viegas:2007wh}, as well as peer production generally~\cite{Shaw:2014gv,Benkler:2002gu}, with a later contribution by Hess and Ostrom analyzing online bioengineering databases~\cite{Hess:2006vx}. 
Others have used the Workshop's resource perspectives to investigate loot distribution norms in the game World of Warcraft~\cite{Ross:2014wc,Strimling:2018vf} and online ``dark institutions'' like software pirate exchanges and hacker collectives~\cite{Sadia:2013ti,Harris:2018db}. There is growing interest in the potential of the Ostrom Workshop to serve digital institution design~\cite{Pitt:2014hr,Pitt:2015bk,Diaconescu:ho,Pitt:2014bp}, but explicitly design-oriented work is rare, and little work has focused on constitutional layers. The constitutional layer defines how a system changes itself and who is formally defined
in the change processes.  For example, as we discuss in more detail below, constitutional layers that are beginning to appear in experiments in cryptocurrency governance focus on issues such as rules for off-chain dispute resolution and processes for adjusting those rules.
Constitutional layers facilitate participatory design similar to how it is practiced in HCI~\cite{muller2003participatory}, extended as an ongoing process. 
A focus on constitutional layers emphasizes that digital institutions can be built with democratic values embedded in them, both in their governance and evolution, not just designed initially in a participatory manner. 

In the following  we  elaborate the normative (participation, accountability, \textit{etc.}) as well as pragmatic (efficiency, resilience, \textit{etc.}) benefits of designing digital institutions.

\section{Designing for change}


A key premise of the democratic Ostrom Workshop perspective is that digital institutions,  in order to be robust to shifting environments, must have policies and processes that define how the institution changes. Critically, these processes must provide mechanisms for ``lower-level'' agents to participate. That is, constitutions providing for stakeholders of several types are necessary for institutions to  endure the challenges endemic to a changing and uncertain environment.


\subsection{Levels of choice in institutions}
Rules are difficult to classify, but successful classification schemes provide vital insights into an institution's structure. 
Rules include statements across several degrees of normativity, from suggestions, to unenforced expectations, to inviolable rules with consequences for noncompliance.  Rules can be formal and informal, they can define and govern many different kinds of units of analysis, and they can apply to complex overlapping subsets of participants.  Under a Workshop framework for classifying rules in terms of the level at which they operate, constitution-level rules are is one of three types:
\begin{itemize}
    \item \textit{Operational rules} concern the lowest-level, most mundane behaviors taken by system members, as constrained by its collective choice processes. On the online marketplace Amazon, operational rules define the elementary actions that each type of user can perform 
    \item \textit{Collective rules} concern the behaviors that the institution performs through the agents authorized to represent it, and functions mostly to define affordances at the operational level. On Amazon they define the market context within which agents operate, and through which collective action processes drive the price mechanism.
    \item \textit{Constitutional rules} concern the space of actions by which the collective choice level is changed and, in the broadest sense, the ``meta'' rules by which the system changes itself. Amazon is by no means a democracy, but like any large corporation defines internal research and review processes under which it evolves. 
\end{itemize}

A system without a constitutional level---without formalized change processes---will either not change at all or will be susceptible to  unstructured or informal drift-like change that, by reducing the faithfulness by which its formal rules describe it, weaken an institution to all but the most powerful. 

These levels help define institutional structure in terms of agent capabilities. For example, on Wikipedia, there are operational rules determining how edits should be performed and evaluated, there are collective action rules defining how conflicts should be resolved, and there are constitutional rules outlining the encyclopedia's extensive body of policy for specifying how policy changes.





\subsection{The need for constitutional layers}

Elinor Ostrom's early thought explicitly ties  institutional self-modification to  cybernetic theories of system fit  and responsiveness to the outside environment~\cite{Ostrom:1995vn}. In the general terms first offered by W. Ross Ashby, a system must be able to change to match the structure and change of its environment~\cite{Ashby:1958ty}. The idea of constitution-layer rule-making articulates this in the context of policy with explicit arguments, taken from Ostrom's famous \textit{Design Principles for Community-based Natural Resource Management}~\cite{Ostrom:1990ws,Cox:2010tb}, that successful self-governing institutions are able to maintain environmental fit, maintain structure at several scales, and include the full range of stakeholders in decision making and meta-decision making.

Defining institutional design so that it includes a constitutional layer has the side effect of expanding the domain of the theory to include participatory systems in which important governance-related decision-making includes more agents than the platform creator or its appointed moderators. 
We argue that, for an institution to be participatory in a meaningful sense, it must provide all agents some avenue for constitution-level action. Ticket systems, polls, and other operational feedback schemes---advocated by Kraut and Resnick, and proposed as much to help administrators maintain legitimacy as to sincerely solicit feedback on their designs---do not faithfully implement participation  unless they are provided an explicit formalized role in constitutional change processes. The system is most participatory that endows all agents with unmediated access to constitution-level choice. A related benefit of a constitutional layer of policy is that it provides a natural place for designers to express values and commit to ethics.


With these considerations in mind, the central arguments of our work are that digital institutions should explicitly define their change processes, that they should define those processes to give several types of stakeholder a stake in meta decisions, and that major problems faced by prominent digital institutions are due to their failures to enact these recommendations.

\begin{table}[t]
\centering
\footnotesize
\begin{tabular}{l|l|l|l|l|l}
\toprule
\textbf{Case} & \textbf{System Architecture} & \textbf{System Environment} & \textbf{Domain} & \textbf{Participants} & \textbf{Dominant Location} \\ \midrule
\textit{Cryptocurrencies}           & Decentralized                & Online                      & Economic        & Mixed                 & International              \\ \hline
\textit{METRC}       & Centralized                  & Physical                    & Economic        & Professionals         & Colorado, United States    \\ \hline
\textit{Minecraft}                  & Mixed                        & Online                      & Social          & Amateurs              & International, United States              \\ 
\bottomrule
\end{tabular}
\caption{A table showing salient dimensions along which our cases vary.  By choosing a varied set of cases we emphasize the relevance of constitutional layers to many different types of digital institutions.}
\end{table}

\section{Case Studies}

In the remainder of this paper, we set out to illustrate the importance of constitutional rules through three case studies of digital institutions.  In the first, we see how cryptocurrencies developed in the anarcho-capitalist tradition have struggled with protocols that are incapable of changing themselves.  In our second case, we see ad hoc constitution-level participatory change in an unwieldy tool for monitoring compliance with cannabis regulations.  In our final case we see a relatively successful market implementation of ``participation-like'' governance in the ecosystem of amateur Minecraft servers.  We selected these cases based on the areas of expertise of the authors of the present paper, and the identification of common conceptual ground among these disparate institutional forms.  The cases are unified by virtue of all being contemporary digitally mediated institutions, but vary along other dimensions (centralized versus distributed software, physical versus online participation, economic versus social domain, professional versus amateur participants).  We avoid standard cases such as Facebook or Twitter to highlight how ubiquitous and varied different forms of digital institutions are, to avoid further contributing to those companies dominating conversations about the design and regulation of digital institution, and to demonstrate the vast potential of constitutional design frameworks to benefit digital society.

\subsection{Case Study 1: Cryptocurrency governance}

\begin{figure}[t]
    \centering
    \includegraphics[width=0.8\linewidth]{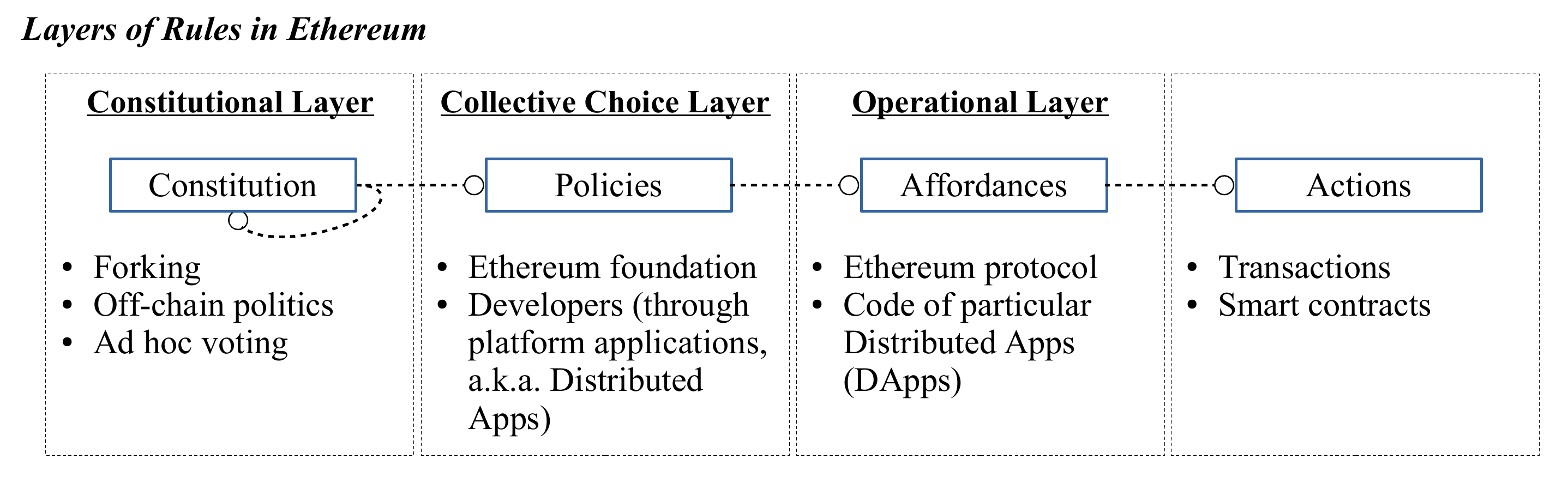}
    \caption{Examples of layers of rules in the cryptocurrency Ethereum.  Constitutional mechanisms are not developed explicitly in Ethereum, so the processes of change in collective choice and operational rules are coarse and chaotic.}
    \label{fig:crypto-summary}
\end{figure}

Cryptocurrencies like Bitcoin and Ethereum are at the frontier of digital institution design, and governance is a central part of the cryptocurrency conversation \cite{bohme2015bitcoin,atzori2017blockchain,dupont2019crypto}. A cryptocurrency is a digital system that maintains a distributed ledger of transactions for the purpose of implementing the functionality of a currency through cryptographic protocols rather than a centralized authority. Cryptocurrency advocates imagine replacing ``fiat'' currencies (such as dollars, euros, and yen) with digital, decentralized  alternatives. 
Cryptocurrencies face the institutional challenges that confront any monetary system, along with the additional challenges of governing a digital decentralized largely anonymous system. We review two design challenges confronting this emerging class of digital institution, 
the ``scaling debate'' and the idea of a proof of stake. The way these issues have unfolded challenges the libertarian and anarcho-capitalist narrative of trustless mechanisms  operating  without institutional scaffolding and constitutional mechanisms beyond market incentives, a narrative that has been enabled by the engineering approach to digital institution design.   
After introducing governance problems typical of today's cryptocurrencies, we discuss the role that constitutional-level rulemaking, or the lack thereof, plays in perpetuating these problems. 




\subsubsection{Constitutional crisis}


Fundamental questions of cryptocurrency standards have in important cases been left to the sociopolitical rather than the market realm. Since the sociopolitical realm has in many cases been assumed away by cryptocurrency designers, the systems they engineer are resistant to adaptive changes, whether minor and prescient or major and existential. 
Two representative design challenges are Bitcoin's scaling debate and the debates around proof of stake.  Difficulties around standards of within-coin evolution demonstrate the need for institutions around each coin that, in many ways, undermine the trustlessness that is supposed to distinguish cryptographic currencies from fiat currencies \cite{werbach2018blockchain}.  

The scaling debate centers around the size of Bitcoin blocks.  Each Bitcoin block is a collection of transactions, and the total number of transactions per block is limited by the size of the blocks. The current maximum allowed block size is just a few megabytes.  As several simultaneous transactions are proposed, their order of execution is computed according to a bidding system, and Bitcoin miners' profits from these auctions are drawn from these bids.  Miners are agents in the currency ecosystem who perform the costly work behind adding each block of transactions to the blockchain, and so they have a profit incentive to keep block sizes small, because that keeps transaction fees high. However, small block sizes are bad for buyers and sellers because new blocks are only created on the blockchain on average every 10 minutes.  If many people are trying to transact,  congestion of transactions in the blockchain can quickly occur, which in turn leads to people being unable to get transactions processed, potentially for days or more. These conflicting incentives between consumers and miners have created an enormous debate around this issue, and  miners' successes at keeping the block size small have severely reduced Bitcoin's competitiveness in the face of newer and more owner-friendly  currencies.  
The Bitcoin scaling debate highlights how multiple actors with different incentives in the cryptocurrency ecosystem create institutional tensions.


A second design challenge faced in cryptocurrency centered around how to securely verify blocks.
Most contemporary coins operate with a ``proof of work'' model in which miners  allocate computing cycles to find hashes that satisfy a criterion.  This mechanism was originally argued to be fair because anyone with a commodity laptop can be a miner and contribute to and benefit from the currency. In practice, the escalating costs of energy and the emergence of specialized hardware have dramatically reduced the profitability of mining to all but a handful of central agents. The people who now have the ability to become profitable miners are those who have the existing capital and mobility that allows them to establish mining stations in places with relatively cheap land and electricity. These problems have driven the development of  ``proof of stake'' mechanisms, which use monetary bids as an alternative to computing cycles. The idea of proof of stake is that an ``ante'' buys you into a block.  If you cheat by creating a false block, you risk losing the money you have bid.  Proof of stake relies on this financial incentive for the correctness of the protocol.  The debate around proof of stake is what the minimum bid should be.  If the bid is too high, fewer people will have the resources to participate in validating blocks, which undermines the vision of decentralization. If the bid is too low, people have no incentive to be honest.   

In both, a finite valuable resource strains agents, but the system prevents them from effectively organizing at the collective level to implement system changes. These problems should not be insurmountable, but a consequence of trustlessness is that currencies are essentially immutable after release: the code supporting a coin has no constitutional layer.  Each change to a currency's protocol, no matter how minor,  requires  a ``fork'' of an existing currency, in which the code from the old coin is copied, edited, and released as a new coin. Holders of the old coin are then all encouraged to simultaneously divest from the old into the new. If a coin's community cannot reach consensus on the necessity of a fork, it bifurcates, and the coin has failed to fulfill the primary function of money: to provide a standard unit of exchange.

Absent formal constitutional mechanisms, what has it looked like in practice? In 2016, a hacker diverted millions of USD from The DAO (Decentralized Autonomous Organization) project to an unknown agent's account. In response, after weeks of debate, the Ethereum Foundation actually called ether coin holders for a kind of vote to undo the theft's damage by  ``reversing time'' in the blockchain ledger of Ethereum through a ``hard fork'' of the system~\cite{Wilcke_forknotfork_2016,dupont2017experiments}. In other words, the response to the crisis of the hack was an ad hoc coordinated voting-with-your-feet process that occurred in large part outside the technical infrastructure of Ethereum, and was in fact coordinated in part by an intervening central authority.  Facing competing pressures from different parts of the community, such as again with miners holding a large amount of power, the Foundation ultimately made the constitution-level decision to put the question to a kind of vote among coin holders.
Throughout the debate around this issue, a hard-line techno-libertarian wing opposed the hard fork on the grounds that the code (even the compromised code) of the system should be treated as the ultimate authority. In an effort to rebut those concerns, community leaders offered reassurance by asserting that voting would be ineffective for resolving more mundane, but equally important policy issues (such as properly parameterizing block size and proof of stake): ``Imagine how hard it would be to get a patch approved, pushed out to mining pools and to get them to reach consensus about a less clear-cut issue. It's just not happening in most circumstances'' \cite{vessenes2016point}.  While cryptocurrencies have shown an ability to change in extreme circumstances, their reliance on the heavy-handed forking mechanism for even the most mundane design decisions translates to a functional lack of constitution-layer rules. Without ``meta'' policies for making granular course-corrections to existing policies, cryptocurrencies are unable to track environmental change, overcome technical or sociotechnical system failures, or keep up with innovations introduced by competing currencies.

\subsubsection{Comparing theoretical guidance}

What are the aspects of the structure of cryptocurrencies as digital institutions that are creating the conditions for the long, grueling debates that have haunted blockchain communities?  How might the engineering tradition analyze these issues, and how might the participatory? The behavior engineering tradition focuses on nudges or incentives as solutions. An incentive-based analysis would suggest incentivizing stakeholders to converge on a block sizes or minimum bid. But the status quo in each case is itself is an unexpected outcome of perverse incentives put in place by the engineering mindset, the same mindset that designed these systems to be immutable. 

In contrast, Ostrom's participatory lens allows us to diagnose a larger issue at play.  This perspective encourages us to not just focus on each of these problems with cryptocurrencies individually, but instead to recognize them as connected manifestations of an underlying lack of rules for resolving debate and making amendments.  Because there are not clear cryptocurrency protocols for  resolving disputes around cryptocurrency protocols, questions around these issues end up yielding to open-ended debates, ad hoc damage control, extralegal maneuvers, and other power plays.  

Our analysis has revealed how several different types of users and stakeholders within the cryptocurrency ecosystem have differing degrees of access to levers of institutional power. A constitution does not only specify change processes, it can also lay out the values that a cryptocurrency is meant to represent, provide a basis for arguments to justify specific institutional changes, and define the specific powers of particular actors and institutional roles.  Institutional structures for discussing and facilitating change can be formulated in a constitutional layer. Without robust systems for resolving debates, currencies increase their exposure to collective action problems that endanger all stakeholders.

From the perspective of the democratic tradition of digital institution design, cryptocurrency communities should accept that it is inevitable that human trust or human power will have a role in complementing the shortcomings of cryptocurrency protocols as monetary institutions \cite{lustig2015algorithmic,atzori2017blockchain}.
The DAO symbolized an interest within the community of embedding governance structures within the protocols of cryptocurrencies themselves, having code functioning as law \cite{dupont2017experiments}. However, these efforts at governance have proven to not go far enough.  Mechanisms of change could in part be implemented in code, but processes for facilitating human debate and interaction regarding these protocols must not be ignored.  ``Off-chain'' politicking will never cease to exist in a system that is robust enough to thrive, and therefore must be considered subject to potential regulation. Moreover, these off-chain debates---whether they are public or clandestine---are best resolved through processes with finer granularity than wholesale acceptance or rejection of entire coin protocols.  

Of course, there is another way that cryptocurrencies can be seen as implementing a constitutional layer. From above the perspective of any individual currency, at the level of the ecosystem of competing currencies, the libertarian implementation of constitutional rulemaking is the market for currencies: currencies compete for coin holders, and coin holders vote for the best protocol ``with their feet'' by divesting from undesirable coins and investing in effective ones, such as occurred after The DAO hack. As technologies advance, and lessons are learned, new coins replace the old ones, and, at the ecosystem level, the change that constitutions provide occurs naturally. This is the key claim of anarchocapitalist thought: the market can implement the key functions of government. Following the democratic tradition, we dispute this claim.   It is unlikely that ``meta-market'' pressures are sufficient to implement or guarantee constitutional-layer rules about rulemaking. Competition may be sufficient to find these optima quickly and efficiently. But special features like high switching costs and network externalities impose formidable inefficiencies, posing a serious obstacle to the effectiveness with which  market competition can implement constitutional change in the ecosystem of currencies  \cite{Cennamo:2013vv,Luther:2016ua,Nair:2017dl}.  Cryptocurrency developers must not just write code, but also detail the principles, values, and rules of off-chain political process.  These considerations are now at the forefront of recent cryptocurrency and blockchain research, and experiments in constitutions are occurring (e.g., \cite{iles2018civil,eos2018}).   Based on the insights of the Ostrom Workshop, our perspective is that those constitutions which prioritize participatory change and afford adaptation to local contexts will be most successful.

\subsection{Case Study 2: Cannabis informatics}

\begin{figure}
    \centering
    \includegraphics[width=0.8\linewidth]{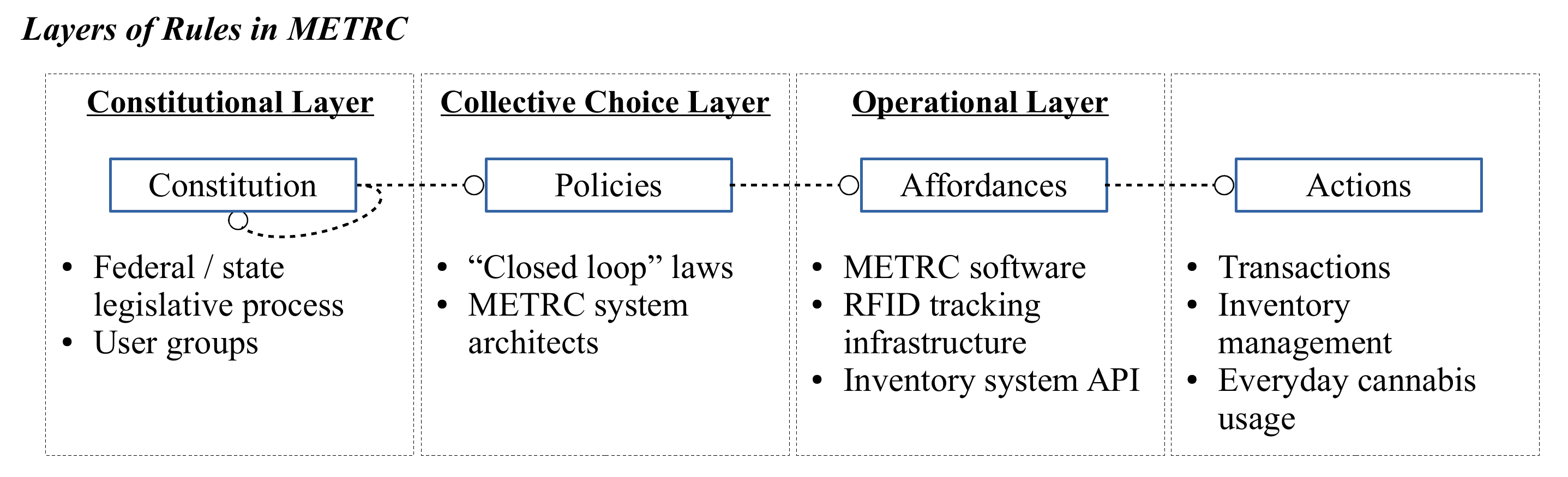}
    \caption{Examples of layers of rules in the METRC cannabis informatics system.  Overarching change processes involve surrounding U.S. legal infrastructure and guidance from user groups. The introduction of user groups provided a bottom-up constitutional mechanism that allowed METRC to address many unanticipated problems.}
    \label{fig:metrc-summary}
\end{figure}

We examine the digital infrastructure involved in the legalized cannabis trade within the United States. Recognition of the medicinal benefits, economic potential, and racial disparities in law enforcement have all contributed to a dramatic shift in attitudes and policy towards legalizing cannabis~\cite{Geiger_sixintenAmericanssupport_2018,schnabel2017should}. Information technologies are playing a central role in the regulation of emerging recreational cannabis markets. While states in the U.S. such as Colorado and Washington have operated regulated marketplaces for recreational cannabis since 2014, the production, distribution, and consumption of cannabis remain federal crimes that carry significant penalties. The legal rationale against federal intervention in these state markets is the ``closed loop theory'' that requires demonstrating cannabis is neither being diverted into the black market nor  crossing state lines~\cite{Houser_HighTimesHistory_2014,Chemerinsky_IntroductionMarijuanaLaws_2017}. A central component of these new legalized markets are government-controlled ``seed-to-sale'' inventory tracking systems that collect and store detailed data about each transaction in the supply chain from planting to retail sale. The creation of an entirely new market commodity, recreational cannabis, at a time of vast digitization has produced a peculiar institution, an ideal case for the place of participation in digital institution design. The history of this system, and its responses to market and cultural shifts, demonstrates the importance of designing digital institutions with adaptability---and wide-ranging participation---in mind.

The Marijuana Enforcement Tracking Reporting Compliance (METRC) system plays a central role in Colorado's \$1.5 billion cannabis industry. METRC is a ``regulatory compliance system'' licensed by the State of Colorado's Marijuana Enforcement Division (MED) from Franwell, a privately-held, supply chain technology services company that offers ``internet-deployed applications and services'' like RFID technologies, originally to improve ``track and trace visibility'' for perishable foods and pharmaceuticals. It was thus generalized from a tool for monitoring single supply chains, to one for monitoring every supply chain of every agent in an entire industry. METRC's implementation by the first states has increasingly made it the technical standard for all subsequent states' legalization efforts.

All medical and retail marijuana businesses in Colorado are required ``to use METRC as the primary inventory tracking system of record''. METRC was built ``by regulators specifically for oversight'' to provide ``the necessary visibility for adherence to rules, regulations and statutes'' to fulfill the legal demands of the ``closed loop theory''~\cite{metrc_system}. METRC has two sides: an industry side that is ``used to report the required events and information'' and a regulatory side ``used for enforcement and compliance monitoring''. The industry side consists of RFID-enabled tags and barcodes attached to every plant and  derivative product that exists, for tracking  ``from seed to sale''. Regulatory compliance begins with tagging seedlings and clones, vegetating and flowering plants, trimmings, business-to-business transfers and distribution, and final sale at a retail location. The inventory and status is entered with specialized tag scanners and web user interfaces into a central database. METRC allows regulators to perform live audits to trace the provenance of every single product in the state-wide marketplace using METRC's own reporting tools within itself, or by using METRC's APIs to pass its data to secondary and tertiary data analysis systems.

\subsubsection{Constitutional crisis}

By reducing compliance to a problem of maintaining provenance in state, the designers of this digital institution were able to assume that  regulating the recreational cannabis market was  isomorphic with the supply chain management problems faced by manufacturers of perishable foods or pharmaceuticals. Fit to local conditions is one of Ostrom principles of successful self-organizing resource management institutions, but this  assumption drove a wedge between the capabilities of the system and the unique demands of a recreational cannabis markets. Another Ostrom principle of successful resdource management is fit to the needs of stakeholders.   METRC did not account for the markets many types of stakeholder  (\textit{e.g.}, growers, distributors, producers, testing facilities, retailers,  \textit{etc.}), the unusual regulatory constraints (\textit{e.g.}, traditional testing facilities like professional laboratories and research universities withheld their services to avoid jeopardizing their federal accreditations), and alternative business models specific to the cannabis industry's needs (\textit{e.g.}, the inability to access FDIC-regulated banks or SEC-regulated capital markets) for a minimally-functional digital institution. 

Consequently, METRC was initially ill-prepared to fulfill the market's demands and there was an avalanche of problems~\cite{subritzky_issues_2016}: the regulatory agency mismanaged the rollout of a predecessor system for medical cannabis~\cite{gorski_pipedream_2013}, there were bottlenecks in manufacturing and distributing enough RFID tags for growers and producers~\cite{ingold_tracking_2013}, the central database had poor usability and uptime that prevented users from uploading their inventories as required by law~\cite{gorski_oversight_2014}, and innovative new products and business models did not fit into METRC's models for logging transactions or ontologies for classifying products~\cite{carnevale_framework_2017,yates_regulation_2018}. As METRC was introduced to regulate a legally precarious marketplace, its fumbled launch jeopardized the larger institutional experiment of legalizing recreational cannabis in Colorado. At the core of this crisis were decisions about the design of a new digital institution and the absence of mechanisms for formally including stakeholders on the ground into change processes.

In a remarkable turn for government information technology deployment,  MED was eventually confronted with the scope and stakes of its failures, and committed to a participatory process for revising the design of the system following its launch. Growers, retailers, regulators, and technologists participated in a series of ``User Group'' meetings to identify the fault lines between the legislatively-mandated affordances, rules-in-use, and emerging practices in the market. A significant development was the introduction of a more robust API that allowed METRC to fulfill its statutory obligations to be a central inventory tracking system while also providing producers and retailers greater flexibility to develop alternative approaches for entering, representing, and retrieving data with secondary and tertiary systems. At present, this API  has end-points for employment, facilities, harvest, items, lab tests, packages, patients, plant batches, plants, rooms, sales, strains, and transfers: an ontological panopoly not recognizably akin to within-firm supply chains for perishable foods or pharmaceuticals, but necessary for the management of a complex ecosystem of stakeholders and resources.  METRC stabilized following the development of these user group feedback mechanisms, enough that  it has become the regulator-mandated seed-to-sale tracking platform in twelve other states' legal medical or recreational cannabis markets.

\subsubsection{Comparing theoretical guidance}
METRC's crisis at launch can be traced back to decisions to design the digital component of this institution with little input from its users. The top-down design approach focused on the superficial similarities between the products moving through cannabis, agricultural, and pharmaceutical supply chains but overlooked how this inventory tracking system migrated into a profoundly different institutional context. In traditional supply chains, actors can enforce compliance with their rules and practices through legal contracts, technical standards, professional norms, and prices. These instruments typically perform the functions that the Ostrom Workshoip identifies with successful governing: defining boundaries and rights, prescribing mechanisms for monitoring and sanctioning, and organizing information around the system. However, METRC's state-mandated inventory tracking system was the bureaucratic implementation of a legislative directive lacking many of these feedback instruments, but compliance could nevertheless be legally enforced. In the absence of substantive feedback mechanisms, the technical system and its bureaucratic managers could not anticipate demand for tags, sustain uptime, or accommodate new business models and product categories. That is was mandated by a democratically governed state did not redeem it, as a legislature representing millions of citizens necessarily operates with a time lag much larger than the emerging market required for policy adjustments.

It is important to emphasize that METRC is not the kind of online community envisioned by Kraut and Resnick; METRC is a government-operated database that collects the daily inventory and transactions of every licensed cannabis operator across Colorado. But seen through the lens of digital institution design, METRC was confronted with  challenges recognizable in Kraut and Resnick's work on building successful online communities, such as cold-starting a new system, socializing newcomers, and regulating behavior. In the case of starting a new system, as a state-sanctioned monopoly METRC did not face the same challenges of carving out a niche, defending the niche from competitors, or building a critical mass of participation. But this institution faced similar design choices around defining the scope of the institution, compatibility with other systems, and organizing information and interactions~\cite[p.232--233]{kraut_building_book_2012}.

Much of the guidance Kraut and Resnick offer comes from an assumption that participation in a digital institution is voluntary. \textit{BSOC} design claims like ``People are more likely to comply with requests the more they like the requester''~\cite[p.32]{kraut_building_book_2012} or ``Face-saving ways to correct norm violations increases compliance''~\cite[p.153]{kraut_building_book_2012} are deeply distorted by non-voluntary institutional contexts like METRC. Compliance with requests from the MED through systems like METRC is incentivized through escalating sanctions like coercion, confiscation, and imprisonment rather than charisma, motivation, or commitment. METRC is not an isolated case: institutions like Transportation Security Administration watch lists, financial and social credit scores, and mobile location traces share similar features of less-than-voluntary participation. How can digital institutions with less-than-voluntary membership be designed to be more accountable, successful, or sustainable? 

The ``choice architecture,'' from the ``nudge'' paradigm of engineering for compliant behavior, would suggest that technocrats who fully apprehend the decision space could provide better defaults, reduce overload, and improved comparability for agents to make ideal decisions. But the decision space of newly legal cannabis markets could not be apprehended because there were no precedents, little information, and complex interactions; there had never been a legal recreational cannabis market; there were no empirical priors from which to draw institutional design guidance. A complex regulatory apparatus governing a legally precarious market with a complex set of incentives and stakeholders nevertheless shows the importance of participatory mechanisms for feedback and institutional design. 

\subsection{Case Study 3: Game server governance}

\begin{figure}
    \centering
    \includegraphics[width=0.8\linewidth]{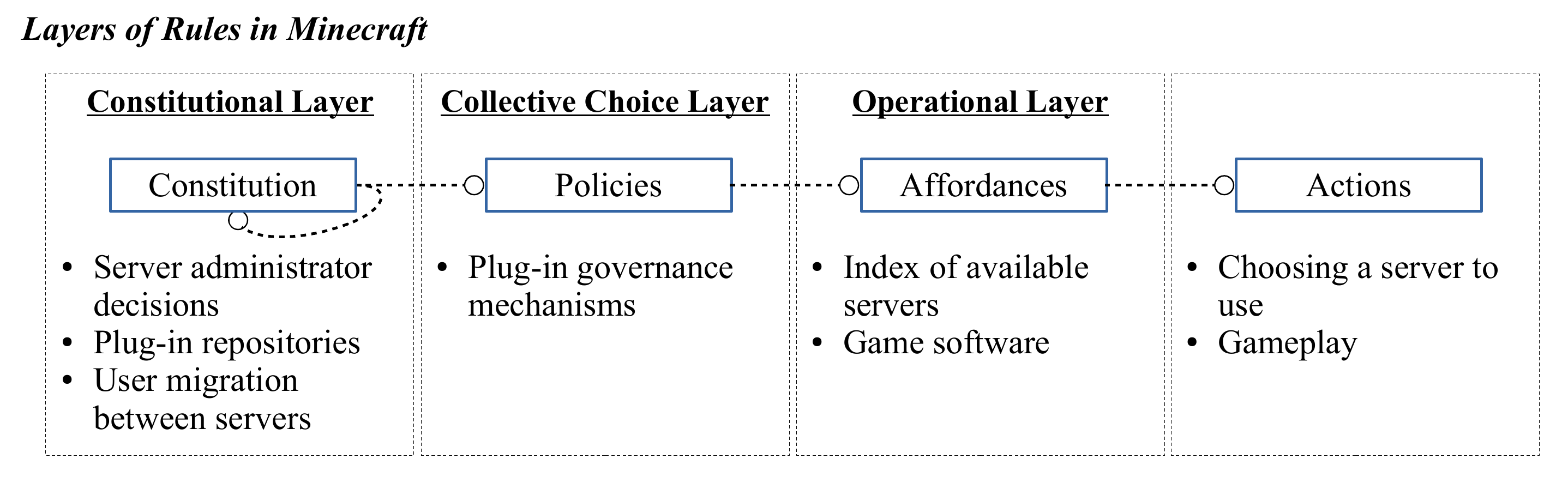}
    \caption{Examples of layers of rules in the game Minecraft. Server administrators can choose from a suite of plugins that facilitate choices in server governance, such as rules of in-game markets or chat capabilities. The rich landscape of plugins at the constitutional layer provides a thriving ecosystem of varied and successful communities.}
    \label{fig:minecraft-summary}
\end{figure}


Governance in digital games is another important area of research \cite{kou2014governance,boellstorff2015coming,crenshaw2017something}.
Minecraft is an open-world (``sandbox'') construction game that is notable for a  culture of amateur-run self-hosted multiplayer servers that players shop between in search of engaging community. Servers must compete with each other for users while struggling to manage several resource provisioning problems, including the needs to manage endemic vandalism by the game's relatively young player base, to provide sufficient RAM, CPU, and network bandwidth for players, and to solve in-game resource problems. Preventing vandalism may mean controlling the availability or effects of in-game phenomena like fire, magma, and TNT that the majority of players enjoy and use responsibly, but that a minority abuse. Providing sufficient physical resources is surprisingly difficult: each additional player to this resource-intensive, lag-sensitive game increases the recommended server requirements by 1GB of RAM, 1--3 Mbits/s of up- and down- bandwidth, and 5GB of cache.  A failure to adequately provide any of these complex sociotechnical services can lead within minutes to users abandoning a server for competing hosts.  Contrary to mythologies about the ``cloud'' abolishing marginal cost economics, Minecraft provides a case where finite and exhaustable digital resources implies the need for robust resource management strategies. 

Fortunately, administrators can address these forbidding governance challenges by drawing upon several tools developed by Minecraft's large informal developer ecosystem. There is a large collection of plugins that automate the deployment and maintenance of rules for peer monitoring, resource monitoring, rule enforcement, trade, vandalism, decision-making, information transmission, and communication, as well as plugins that define complete market and property rights institutions: systems of property rights, shops, social hierarchies, and group allegiances. The richness and granularity of plugin governance modifications has reduced many of the challenges of amateur governance to the flipping of switches, and has made possible an exciting diversity of governance styles.  The library of governance plugins available to an administrator leaves it entirely to their discretion what governance capabilities their server will have.  
Administrators seem to take full advantage of this flexibility. Among the diverse amateur digital institutions we have observed, and well beyond the professionalized for-profit servers which exist in their own ecosystem, are a groups of friends who install no rules but punish infractions by chiding each other in person, a pristine server that solves its key resource problem by preventing visitors from changing anything, and an explicitly anarcho-capitalist server with privatized law enforcement system, in which users secure justice for themselves by placing rich bounties for the virtual heads of those that have wronged them. There are active, long-lived communities with as few as 2 users and as many as 2,000. Administrators are effectively implementing their na{\"i}ve folk theories of cooperation, incentive design, and resource management, and iterating over them. The lessons learned by these governance autodidacts are then reified into the plugins that are refined, repropagated, and refined again.

\subsubsection{Constitutional crisis}
What governance challenges do Minecraft server operators face, especially small-scale operators with fewer resources and a greater reliance on community members? How should operators  design their digital institutions to manage these resource challenges? And, specifically, what is the role of servers' constitutional choice mechanisms in managing or exacerbating these problems?
Minecraft servers fail when they stop being visited, something that can happen when their administrators fail to manage the many finite resources their server needs to function, and when they fail to provide users with a stable and secure environment safe from vandalism and harassment. A server's sensitivity to these threats increases with server size.  But because they are all relatively small, defined by default by a single administrator, and because they exist entirely in software, individual servers can change themselves easily.  

Of course, the existence of constitutional-level rules and wide-spread participation in those rules are different.  Minecraft servers are not democratic by default. Administrators often make themselves available to user feedback either directly or through ticket or forum systems, but it is rare for these mechanisms to be substantively participatory, or formally defined into the processes of constitution-level change.

Zooming out from the individual server, to view the system of competing servers as a digital institution, another mechanism for democratic involvement in constitutional change becomes clear, one very much like that available in cryptocurrencies.  Because servers compete with each other for users, users can, again, vote with their feet, departing servers that do not explicitly or incidentally represent their needs, and populating or even reproducing  servers that do.  Because of this structure, Minecraft servers fare well as a community.  Collectively these seem able to solve the resource management problems that individual servers so often fail to overcome.

This success raises a question: why does the market approach to constitutional change seem to work in Minecraft but not in cryptocurrency?  The answer may be that, compared to cryptocurrencies, Minecraft servers have much lower switching costs: no need to divest, low value contributions, and less role of coordination. 
With these ingredients in place, Minecraft servers that are individually autocratic can as a collective of competitors implement meaningfully participatory change processes typical of the democratic institutions that the Ostrom Workshop describes.

\subsubsection{Comparing theoretical guidance}
The amateur Minecraft community builder seems especially suited to adopt the  engineering approach to digital institution design. These relatively small-scale online communities have one self-appointed leader with complete control over the structure of their virtual world, a leader  highly incentivized to overcome the obstacles to successful community.  As such, the elements identified by engineering-focused frameworks such as Kraut and Resnick's map directly to many of the problems that Minecraft server administrators face.  Attracting, socializing, and retaining new members, and motivating rule compliance all describe problems areas faced by amateur Minecraft administrators, and the engineering approach offers solutions to these problems that administrators can benefit from directly, because they have the power to redesign the system in a way that implements ``best practices'' extracted from previous social engineers who have faced the same problems. 

However, the behavioral manipulations of the engineering paradigm misses important dimensions of governing a self-hosted game server, and important mechanisms for ensuring its success.  For example, Kraut and Resnick's cases tend to  consider  communities in isolation, largely ignoring the fact of competing communities pursuing the same work in their own unique ways, diverging along some dimensions of governance and culture, and converging along others. By failing to give servers' ecological context a central role in the framework, they miss the strategic dimension that competition adds to an administrator's reasoning, and the strong incentives administrators have to make decisions in the community's interest. The Ostrom Workshop, by contrast, is as much influenced by ecology as economics, and has several multi-scale frameworks for incorporating an institution's social ecological contest into the analysis.

With its view of governance largely focused on resource governance, the Ostrom Workshop's design principles are a valuable guide for the type of governance challenge facing a self-hosted game server. 
Of most relevance for our analysis, which has focused on the  mechanisms that digital institutions have to change themselves, are those principles that recognize the value of conflict as a symptom of the need for change, those that encourage institutional fit to the environment and, most important, that that recommend a role for users in how a digital institution changes.

\section{Discussion}
The participatory tradition in digital institution design, kicked off in large part by Kollock and Smith~\cite{kollock_managing_1996}, has lagged behind other more convenient paradigms, most notably the behavior engineering tradition, which specifies a community around the leveraging of behavioral regularities. The cost of convenience has been a pandemic of user abuse with staggering consequences both online and in the real world. This abuse has been both by system designers, often placing profits above users' basic rights to privacy, and by outside political agents manipulating platforms toward malicious ends.  The result is not merely that major digital platforms misuse their users, but, from a larger perspective, that they fail to keep up with users' needs or the constraints of a sociotechnical environment in flux.  We argue that designing digital institutions for change, including bottom-up change, makes them more humane and adaptive. We articulate this in terms of the need for attention to an institution's constitution-level rules.  In support of our argument, we review the influence that the work of Elinor Ostrom has had on participatory approaches to digital institution design, and we apply the taxonomy of rule levels to inspect three digital institutions, the challenges they face, and the role of formal participatory change processes, or lack thereof, in each institution's problems. 

Our analysis of the cryptocurrency ecosystem belies its ideological commitment to purely technical (rather than sociotechnical) governance, and shows that the community's attempt to remove humans from the operation of human institutions has produced currencies that are inflexible. In this ecosystem, each currency's inevitable need for incremental policy changes can be met only with blunt policy tools whose side effects undermine the very communities they are intended to serve. 

We then analyze the regulatory ecosystem around recreational cannabis in Colorado, to show how top-down management, incongruent ontologies, and ponderous change processes conspired to produce a digital supply chain surveillance tool that caused a crisis in the industry, one that was only overcome when stakeholders worked outside of established channels to amend the digital regulatory infrastructure to meet user demands.

In our third case, we discuss the organizational structure supporting multiplayer activity in the sandbox video game Minecraft, the prominence of which is due in large part to volunteer amateur administrators self-hosting and learning to govern independent servers, in what amounts to hundreds of thousands of small-scale experiments in self-governance. We find that in order for a server to succeed, its administrator must learn how to foster successful collective action ``on the job'' by formulating policy and installing code modules that implement the basic functions of resource governance.  Where administrators fail to meet the demands of users or their environment, competition between administrators for traffic between their servers leverages market structures to implement ecosystem-wide constitution-level change.  Why ``vote with your feet'' works in Minecraft, where it seems to fail in the cryptocurrency case, seems due to several factors, including the immutable nature of individual currencies, the higher costs in them of exiting or performing most other actions, and an existential difference between the two types of platform, that while game servers may benefit from convergence on one standard community, currencies require that convergence, such that network externalities play a much bigger role in the dynamics of the cryptocurrency ecosystem. 

Through all of these we draw a contrast with the engineering tradition in digital institution design, exemplified by Building Successful Online Communities, whose authors have not only had a tremendous influence on how academics think about digital institutions, but have contributed brainpower to the largest digital institutions today, and steered Facebook, Twitter, and other major actors toward structures that are not accessible or accountable to their users in any intentional way, and are starting to suffer for it.

Overall, we argue that digital institutions must be designed to evolve from the bottom up. The thesis is not especially controversial to technology researchers: it comes to resource economics via cybernetics, whose lasting influence on the science of sociotechnical systems occasionally reveals itself in new ways.  Still, actual frameworks and theory for understanding institution change have failed to bring the necessary attention to participatory approaches to digital institution design.  With new tools for bringing democratic values more concretely into the digital institution design, we hope to contribute to more consciously adaptive, value-driven institutions.

\section{Conclusions}

In 1996, Kollock and Smith offered a prescient conclusion to their groundbreaking work, ``As computer-mediated communication increasingly becomes the media through which public discourse takes place, the ways in which that discourse is socially organized becomes more consequential.''  

In the present work we have argued that existing design processes have excluded \textit{individual users from substantive roles in system change}, preferring  instead to embed them into lower-level roles , and  interfaces engineered to nudge behavior to suit  design goals.  The fields of human computer interaction and computer-supported cooperative work find their history in the design of interfaces to support individual user experiences and the shared experiences of small groups of users.  As these research communities have moved their attention to larger-scale settings, they have often brought the same approaches and limited image of the user's role.  While useful for certain purposes, the engineering tradition of institution design fails to incorporate values fundamental to fair, adaptable, and ethical digital institutions.

Political scientists, economists, and ecologists in natural resource governance have spent decades building the argument that the most effective management of local environments is often through democratically designed and managed institutions \cite{Ostrom:1990ws,ostrom1995incentives}.  These communities are producing growing quantitative evidence that institutional structures that are not tailored to local contexts and do not include mechanisms of local stakeholder-driven change are vulnerable.  However, these methods and frameworks are only slowly being applied to digital institutions. In our case studies, we have shown how these problems manifest in three highly varied digital institutions.  Constitutional layers of digital institutions are necessary to facilitate democratic processes of social change.  Without meaningful design for inclusive change, our digital institutions are susceptible to drift amorally toward abusive dynamics, and users are vulnerable to abuse. 


\section{Acknowledgments}
Omitted for review.

\bibliographystyle{plain}
\bibliography{manuscript}

\end{document}